\documentclass[12pt]{article}
\advance\topmargin by -1.6cm

\newenvironment{eq}
{\[\begin{array}}{\end{array}\]{}}

\let\rvec=\vec        

   \def\({\Bigl(}
\def\){\Bigr)}    \def\|{\Big|}
\def\then{~\Rightarrow~}   \def\o{\circ}
    \def\x{\times}
   \def\ox{\otimes}
 
\def\pl{{~\oplus~}}

\def\PL{\displaystyle \bigoplus}
\def\SUM{\displaystyle \sum}

\def\mid{\big\bracevert}

\def\sub{\subseteq}
\def\subnoteq{\subset}

\def\supnoteq{\supset}
\def\and{\wedge}

\def\AND{\displaystyle\bigwedge}

\def\OD{\displaystyle\bigvee}

\def\rin{{\,\in\kern-.42em\in}}

\def\tr{{\,{\rm tr }\,}}
\def\det{\,{\rm det }\,}

\def\centr{\,{\rm centr}\,}

\def\weights{{{\bf weights}\,}}
\def\irrep{{{\bf irrep\,}}}

\def\A{{\,{\rm A\kern-.55emA}}}
\def\B{{\,{\rm I\kern-.2emB}}}
\def\C{{\,{\rm I\kern-.55emC}}}
\def\E{{\,{\rm I\kern-.2emE}}}
\def\G{{\,{\rm I\kern-.55emG}}}
\def\H{{{\rm I\kern-.2emH}}}
\def\I{{\,{\rm I\kern-.2emI}}}
\def\K{{\,{\rm I\kern-.2emK}}}
\def\L{{\,{\rm I\kern-.2emL}}}
\def\M{{\,{\rm I\kern-.16emM}}}
\def\N{{\,{\rm I\kern-.16emN}}}
\def\Q{{\,{\rm I\kern-.5emQ}}}
\def\R{{{\rm I\kern-.2emR}}}
\def\S{{\,{\rm I\kern-.42emS}}}
\def\T{{\,{\rm I\kern-.37emT}}}
\def\UU{{\,{\rm I\kern-.51emU}}}
\def\Z{{\,{\rm Z\kern-.32emZ}}}

\def\al{\alpha}  \def\be{\beta} \def\ga{\gamma}
\def\de{\delta}  \def\ep{\epsilon}  
    
   \def\la{\lambda}   \def\si{\sigma}
   \def\om{\omega} 
\def\phi{\varphi}

\def\vec#1{\underline{\bf vec}_{#1}}

\def\SL{{\bf SL}}
\def\U{{\bf U}}

\def\SU{{\bf SU}} 
 
\def\SO{{\bf SO}}

\def\d#1{{\check{#1}}}
\def\angle#1{\langle#1\rangle}

\def\bl#1{{\bf {#1}}}
\def\cl#1{{\cal #1}}

\def\lvec#1{\stackrel{\leftarrow}{#1}}

\def\map{\longrightarrow}

\def\mape{\longmapsto}

\begin{document}

\newpage

\begin{titlepage} 
\hfill MPI-PhT/00-46 

\hfill hep-th/

\vskip25mm
\centerline{\bf THE SYMMETRY REDUCTION}
\vskip3mm 
\centerline{\bf FROM INTERACTIONS TO PARTICLES}
\vskip15mm
\centerline{
Heinrich Saller\footnote{\scriptsize 
saller@mppmu.mpg.de}  
}
\centerline{Max-Planck-Institut f\"ur Physik and Astrophysik}
\centerline{Werner-Heisenberg-Institut f\"ur Physik}
\centerline{M\"unchen}
\vskip25mm

\centerline{\bf Abstract}
\vskip10mm
The hypercharge-isospin-color symmetry of the standard 
model interaction is
drastically reduced to a remaining Abelian electromagnetic $\U(1)$-symmetry
for the particles. It is shown that such a symmetry reduction comes as a
consequence of the central correlation in the 
internal group as represented by the standard fields where the hypercharge 
properties are given by the central
isospin-color properties.
A maximal diagonalizable symmetry subgroup (Cartan torus)
of the interaction group for the
particles as eigenvectors has to discard either color (confinement)
or isospin.
An additional diagonalization for the external spin properties which
come centrally correlated with the isospin properties enforces the weak isospin
breakdown.

\vskip5mm

\end{titlepage}

{\small \tableofcontents}

\newpage

\section{Introduction}

The interactions in the standard model\cite{WEIN} of elementary particles
are invariant under the external  transformations
with the semidirect Poincar\'e group $ \R^4\lvec \x\SO_0(1,3)$
(with respect to halfinteger spins 
written with $\SL(\C^2)$ as the twofold 
covering group of the Lorentz group $\SO_0(1,3)$)
and under the internal operation group defining hypercharge, isospin and color
properties 
\begin{eq}{l}
\hbox{\it interaction symmetry: }\underbrace{\R^4\lvec \x\SL(\C^2)}_{{\rm external}}\x
\underbrace{\U(1)\x\SU(2)\x\SU(3)}_{{\rm internal}}
\end{eq}The standard interactions are implemented by the 12 internal gauge fields
which come as 4-vectors with respect to the external Lorentz group.
I shall show below that slight, but important changes should be made 
in this group with respect to the faithfulness of its representation.

There is a dramatic breakdown\footnote{\scriptsize
A blowup of symmetries as a consequence of a linearization (tangent space
expansion) is proposed in \cite{S982}.} 
 from the real $(10+12)$-parametric 
Lie symmetries\cite{FULHAR,HEL,LIE13} for the interaction to the symmetries for the particles 
\begin{eq}{rl}
\hbox{\it massive particle symmetry:}&\underbrace{\R\x \SU(2)}_{{\rm external}}\x
\underbrace{\U(1)}_{{\rm internal}}\cr\cr
\hbox{\it massless particle symmetry:}&\underbrace{\R\x \U(1)}_{{\rm external}}\x
\underbrace{\U(1)}_{{\rm internal}}\cr
\end{eq}where Wigner's definition\cite{WIG} for free particles as unitary irreducible
representations of the Poincar\'e group is used. In this strict sense,
confined quarks are no particles, they do not have 
a mass as  eigenvalue for the spacetime translations.
The Poincar\'e group representations are induced\cite{MACK} 
by  representations of direct product  subgroups which have a
rotation factor - either  $\SU(2)$
with spin numbers $J=0,{1\over2},1,\dots$ or
axial rotation (polarization) $\U(1)$ 
numbers $M=0,\pm1,\pm2,\dots$ for massive and massless particles resp.,
and a time translation factor $\R$  represented in a rest system
with the mass $m$ as eigenvalue 
for massive particles  $m^2=q^2>0$ and
in a polarization system 
with the absolute value of the momentum $|\rvec q|=q_0$
as eigenvalue for massless particles  $m^2=q^2=0$, $|\rvec q|>0$.

The word symmetry - in connection with multiplicity - 
is used in its strict sense: E.g. 
as particles, a proton and a neutron
may be called an isospin induced or isospin related 
doublet, but not an isospin symmetric  doublet - 
with their different masses there is no 
$\SU(2)$-symmetry connecting those two particle states. Or, more obvious,
the four particles comprising the weak bosons $\{Z^0,W^\pm\}$ and
the photon $\ga$
are no isospin symmetric triplet-singlet, 
there is no symmetry transformation left
between  them.

The internal symmetry reduction from 
interaction parametrizing fields to 
assymptotic free particles has two aspects: 
Nontrivial color $\SU(3)$-re\-pre\-sen\-ta\-tions are confined, there are even no
 nontrivial  color induced multiplets seen in the particle regime.
Nontrivial  isospin induced muliplicities remain visible 
 in the case of the hypercharge-isospin breakdown   which
 is asymptotically reduced to an electromagnetic Abelian $\U(1)$-symmetry. 

As familiar from the energy eigenstates of quantum mechanics, 
particles 
are constructed as eigenvectors with 
respect to a maximally diagonalizable
subgroup, including the time translations, with the corresponding weights collecting
the eigenvalues for the operations involved.
E.g. eigenstates for electromagnetic $\U(1)$-operations
are characterized by
integer charge numbers $z$, spin $\SU(2)$-eigenstates with respect to an
$\SO(2)\cong\U(1)$-subgroup (3rd spin direction)
by eigenvalues 
$|J^3|\le J$ for a spin $(2J+1)$-plet
etc. 
The weights of massive particles are given by $(m,J^3,z)\in\R\x\Z\x\Z$
with the components for mass, 3rd spin direction and integer chargelike number
(particle-antiparticle)  
and of massless particles  by the weights $(|\rvec q|,M,z)\in\R\x\Z\x\Z$ 
with the components for
momentum absolute value,  polarization and charge.
Therefore, the transition from the large interaction symmetry group to the small
particle symmetry group has to discuss the problem of
maximal diagonalizable subgroups of the interaction group.

\section{Central Correlations}

An important feature of the operation groups where  eigenvectors are looked for
is their central correlation structure to be explained and 
exemplified in the old example of the quantum mechanical Kepler
potential (hydrogen atom\cite{FOCK}) 
and in the internal standard model interaction\cite{HUCK,S921,RAIF}
symmetry. 

A direct product of two  groups $G_1\x G_2$
becomes centrally correlated by considering  the quotient group defined by the
classes  
with respect to a nontrivial  subgroup $C$ in the centers of both factors
\begin{eq}{l}
{G_1\x G_2\over C},~~\{1\}\ne C\sub\centr G_1\cap\centr G_2
\end{eq}

The following Lie groups will be considered
\begin{eq}{l}
n=1,2,\dots:~~
\centr\U(n)\cong\U(1)\supnoteq\I(n)\cong\centr\SU(n)=\centr\SL(\C^n)\cr

\end{eq}The centrum of $\SU(n)$ can be written additively as
$\Z\hbox{mod} ~n$ or, multiplicatively, as the cyclotomic group
$\I(n)$
\begin{eq}{l}
\Z\hbox{mod} ~n=\{0,1,\dots,n-1\}\cong\I(n)=\{e^{2\pi i k\over n}\mid k=0,\dots,n-1\}
\end{eq}

Groups $\SU(n)$ and $\SU(m)$ with $n$ and $m$ relatively prime (no
common nontrivial divisor), e.g. isospin $\SU(2)$ and color $\SU(3)$
cannot be centrally correlated.

A covering group as Lie algebra exponent gives 
rise to locally isomorphic groups,
i.e. with isomorphic Lie algebras, by classes with respect to
discrete centrum subgroups with the familiar examples
\begin{eq}{l}
\R/\Z\cong\U(1)\cong\U(1)/\I(n)\cr
\hbox{$k$ divisor of $n$: 
}\SU(n)/\I(k)\hbox{ e.g.}\left\{\begin{array}{l}
\SU(2)/\I(2)\cong\SO(3)\cr
\SL(\C^2)/\I(2)\cong \SO_0(1,3)\cr
\SU(3)/\I(3)\cr
\SU(4)/\I(2),\SU(4)/\I(4),\dots\end{array}\right.
\end{eq}

Obviously, the irreducible representations and the weights of a
centrum classified group are subsets of
those for the  unfactorized group.

\subsection{The Eigenstate Squares of the Hydrogen Atom}

The perihel conservation
in the orbits as  solutions of the Kepler Hamiltonian
\begin{eq}{l}
H={\rvec p^2\over 2}-{1\over |\rvec x|}
\end{eq}is described by the Lenz-Runge vector
$\cl F$ which defines a 3-parametric invariance 
in addition to the position rotation 
$\SO(3)$ invariance with the angular momenta $\cl L$
as elements of the rotation Lie algebra\footnote{\scriptsize
$\log G$ denotes the Lie algebra for the Lie group $G$.}
 $\log\SO(3)$
\begin{eq}{l}
\rvec{\cl L}=\rvec x\x\rvec p,~~\rvec {\cl F}=\rvec p\x\rvec{\cl L}-{\rvec
x\over r}\cr 
[\rvec{\cl L},H]=0,~~[\rvec{\cl F},H]=0
\end{eq}As shown by Fock, these invariances 
indicate - not repeating all the subleties found in the literature - 
an interaction symmetry for the bound states 
with energy $E<0$ with the real 6-dimensional Lie algebra
with basis 
$\{\rvec{\cl L}_\pm={1\over2}(\rvec{\cl L}\pm{\rvec{\cl F}\over\sqrt{-2H}})\}$
\begin{eq}{l}
\log[\SU(2)\x\SU(2)]\cong\R^6 
\end{eq}Therewith the bound states are acted upon with
representations of the  direct product two factor group 
$\SU(2)\x\SU(2)$ involved
whose irreducible representations  are
characterized by two integer or halfinteger `spin' numbers
\begin{eq}{rl}  
\irrep[\SU(2)\x\SU(2)]
&=\irrep\SU(2)\x\irrep \SU(2)\cr
\irrep\SU(2)&\cong\{[J]\mid J=0,{1\over2},1,\dots\}\cr
\weights\SU(2)&=\{M\mid M=0,\pm{1\over2},\pm1,\dots\}\cr
\end{eq}

However, the energy-degenerated multiplets experimentally obseved are
all  squares, i.e. characterized by
two equal `spin' numbers for both factors
\begin{eq}{l}
[J;J]=[0;0],[{1\over2};{1\over2}],[1;1],\dots\cr
\hbox{with multiplicities }(2J+1)^2=1,4,9,\dots\cr
\weights[J;J]=\{(M_1,M_2)\mid |M_{1,2}|\le J\}
\end{eq}$2J+1$ is the principal quantum number with
$E=-{1\over 2(2J+1)^2}$ the energy.
For the nonrelativistic
hydrogen theory the internal two spin directions for the electron
leading to the observed  multiplicities $2(2J+1)^2=2,8,18,\dots$ have to 
be taken into account by hand.

The correlation between the two `spin' numbers $[J_1;J_2]$
in the Kepler dynamics  is a consequence
of the orthogonality of angular momentum and
perihel vector
\begin{eq}{l}
\rvec{\cl L}
\rvec{\cl F}=0
\end{eq}This orthogonality induces a central correlation:
The group maximally faithfully represented on the bound states  is
not $\SU(2)\x\SU(2)$, but  a quotient group which correlates the
centers of both factors
\begin{eq}{l}
\centr\SU(2)=\{\pm \bl 1_2\}\cong\I(2)
\end{eq}The equivalence group is the 
`synchronizing' cycle $\I(2)$ in the bicycle $\I(2)\x\I(2)$ (Klein group)
\begin{eq}{l}
\centr[\SU(2)\x\SU(2)]=\{(\pm \bl1_2,\pm\bl1_2)\}
\supnoteq\{(\bl1_2,\bl1_2),(-\bl1_2,-\bl1_2)\}\cong\I(2)
\end{eq}The irreducible representations of the group with the
equivalence classes
\begin{eq}{l}
{\SU(2)\x\SU(2)\over \I(2)}\cong\SO(4),~~\centr\SO(4)=\{\pm\bl1_4\}\cong\I(2)
\end{eq}are characterized by an integer sum
of both `spin' numbers $J_1+J_2\in\N$.
They  come in two types
\begin{eq}{rll}
J_1=J_2=J:&[J;J]&\hbox{with }J=0,{1\over2},1,\dots\cr
J_1\ne J_2:&[J_1;J_2]\pl [J_2;J_1]&\hbox{with }
J_1,J_2=0,1,2,\dots\cr
\end{eq}with the eigenvalues (weights) in the
first  case
either both integer or both halfinteger and, in the 2nd case, 
both integer
\begin{eq}{rl}
\weights\SO(4)_{J_1=J_2}&=\{(M_1,M_2)\mid 2M_{1,2}\in\Z,~M_1+M_2\in\Z\}\cr
\weights\SO(4)_{J_1\ne J_2}&=\weights\SO(3)\x\weights\SO(3)\cr
\weights\SO(3)&=\{M\mid M\in\Z\}\cr
\end{eq}with the defining  representations, faithful and not faithful
for $\SO(4)$ 
\begin{eq}{lrl}
J_1=J_2:& [{1\over2};{1\over2}]:&{\SU(2)\x\SU(2)\over\I(2)}\map\SO(4)\cr
&&(-\bl1_2,\bl1_2)\mape -\bl1_4\cr
J_1\ne J_2:& [1;0]\cong[0;1]:&{\SU(2)\x\SU(2)\over
\I(2)}\map\SO(3)\cong\SU(2)/\I(2)\cr
&&(-\bl1_2,\bl1_2)\mape +\bl1_3\cr
\end{eq}

The orthogonality above enforces even $J_1=J_2$,
i.e. it allows only the complex irreducible representations
where the  weights occupy squares. 

\subsection{The Hypercharge Correlation with Isospin-Color} 

The fields of the standard model transform under isospin $\SU(2)$
with the irreducible representations
\begin{eq}{l}
\irrep\SU(2)=\{[2T]\mid T=0,{1\over2},1,\dots\}\cr
\hbox{multiplicities: }2T+1
\cr 
\end{eq}as well as under color $\SU(3)$ with the irreducible
representations characterized by two integers
\begin{eq}{l}
\irrep\SU(3)=\{[2C_1,2C_2]\mid  C_{1,2}=0,{1\over2},1,\dots\}\cr
\hbox{multiplicities: }(2C_1+1)(2C_2+1)(C_1+C_2+1)
\cr 
\end{eq}

From now on, I use integers, odd and even,
for the weights and representations replacing the
  halfintegers and integers 
  as used for familiarity in the Kepler dynamics above. The integers are the winding numbers 
  $z\in\Z$ characterizing the representations of $\U(1)$-subgroups.

The left handed quark and antiquark isodoublet
color triplet fields are examples for the 
complex 6-dimensional defining dual representations 
of isospin-color
\begin{eq}{l}
\irrep\SU(2)\x\irrep\SU(3)=\{[2T;2C_1,2C_2]\}\cr
\hbox{defining representations: }u=[1;1,0],~~\d u=[1;0,1]
\end{eq}The totally antisymmetric tensor powers of
the defining representations generate - up to isomorphy -
all fundamental representations
for isospin and color  by the products 
\begin{eq}{l}
{\AND^n}u\ox{\AND^m}\d u,~~n,m\in\N\then\left\{\begin{array}{ll}
\hbox{for }\SU(2):&{\AND^3}u\cong[1]\cong {\AND^3}\d u,~~\hbox{doublet}\cr
\hbox{for }\SU(3):&\left\{
\begin{array}{ll}
{\AND^2}u\cong[0,1],&\hbox{antitriplet}\cr
 {\AND^2}\d
 u\cong[1,0],&\hbox{triplet}\cr\end{array}\right.\cr\end{array}\right.
\end{eq}

Therewith the hypercharge number $y$ of the interaction fields
in the standard model is a consequence of their
isospin-color powers
\begin{eq}{l}
6y=n-m
\end{eq}as shown in the table
\begin{eq}{c}
\begin{array}{|c||c|c|c|c|c|c|}\hline
\hbox{\bf field}&\hbox{\bf symbol}&(n,m)&
\U(1)&\SU(2)&\SU(3)\cr
              & &&y={n-m\over6} &[2T]& [2C_1,2C_2]\cr\hline\hline
\hbox{left lepton}&\bl l&(0,3)&-{1\over2}&[1]&[0,0]\cr\hline
\hbox{right lepton}&\bl e&(0,6)&-1&[0]&[0,0]\cr\hline
\hbox{left quark}&\bl q&(1,0)&{1\over6}&[1]&[1,0]\cr\hline
\hbox{right up quark}&\bl u&(4,0)&{2\over3}&[0]&[1,0]\cr\hline
\hbox{right down quark}&\bl d&(0,2)&-{1\over 3}&[0]&[1,0]\cr\hline\hline
\hbox{Higgs}&\bl \Phi&(3,0)&{1\over2}&[1]&[0,0]\cr\hline
\hbox{hypercharge gauge}&\bl A&(0,0)&0&[0]&[0,0]\cr\hline
\hbox{isospin gauge}&\bl B&(1,1)&0&[2]&[0,0]\cr\hline
\hbox{color gauge}&\bl G&(1,1)&0&[0]&[1,1]\cr\hline
\end{array}\cr\cr

\hbox{\bf Hypercharge of  the Standard Model Fields}
\end{eq}

The hypercharge is related to
the two-ality of the $\SU(2)$-re\-pre\-sen\-ta\-tions
and the  triality\cite{BIED} of the $\SU(3)$-re\-pre\-sen\-ta\-tions
by the modulo relations 
\begin{eq}{rrl}
\hbox{isospin two-ality:}&2T\hbox{mod}2&=6y\hbox{mod}2\cr
\hbox{color triality:}&2(C_1-C_2)\hbox{mod}3&=6y\hbox{mod}3\cr
\end{eq}

The centrality ($n$-ality) $k\hbox{mod}~n$ of an 
$\SU(n)$-re\-pre\-sen\-ta\-tion 
describes the centrum representation involved
\begin{eq}{l}
\I(n)\ni e^{2\pi i \over n}\mape e^{2\pi i k \over n}\in\I(n)
\end{eq}e.g. faithful for $\SU(2)$-re\-pre\-sen\-ta\-tions 
with two-ality $2T\hbox{mod}2=1$ and  for $\SU(3)$-re\-pre\-sen\-ta\-tions with
triality $2(C_1-C_2)\hbox{mod}3=\pm1$.

This central correlation shows that the group maximally faithfully
represented by the fields in the standard model 
is given by the following classes of the direct product group
\begin{eq}{l}
{\U(1)\x\SU(2)\x\SU(3)\over \I(6)}
={\U(2)\x\SU(3)\over\I(3)}=
{\SU(2)\x\U(3)\over \I(2)}=\U(2\x3)
\end{eq}The representation of the  
subgroup `synchronizing' both centrums 
$\I(2)\x\I(3)=\I(6)\subnoteq \U(1)\cap[\SU(2)\x\SU(3)]$ 
\begin{eq}{rl}
\I(6)\x\I(6)=&\{
(e^{{2\pi ik_1\over6}},e^{{2\pi ik_2\over6}})\mid k_{1,2}=0,\dots,5\}\cr
\supnoteq 
&\{(e^{{2\pi ik\over6}},e^{{2\pi ik\over6}})\mid k=0,\dots,5\}\cong\I(6)
\end{eq}determines the 
hypercharge numbers as integer multiples of ${1\over 6}$.

The  eigenvalue spectrum for the representations of the centrally
correlated internal group is
\begin{eq}{rl}
\weights \U(2\x3)
=\{[y||2t;2c_1,2c_2]\mid& 2t,2c_{1,2}\in\Z,\cr
&\hbox{with }y\in t-{c_1-c_2\over3}+\Z\}
\end{eq}

\section{Cartan Tori}

A Lie algebra has Cartan subalgebras, for
semisimple Lie algebras given by maximal Abelian 
subalgebras, diagonalizable in a representation. 
Going from a Lie algebra to 
its exponent, a  Lie group,
a Cartan subalgebra gives rise to a Cartan subgroup.
A maximal Abelian direct product subgroup of a compact group
\begin{eq}{l}
\U(1)^n=\underbrace{\U(1)\x\cdots\x\U(1)}_{n-{\rm times}} 
\end{eq}will be called an {\it $n$-dimensional Cartan torus} which may be parametrized
for each direct factor (`circle') by  
\begin{eq}{l}
\U(1)=\{e^{i\al}\mid\al\in[0,2\pi[\}
\end{eq}

If  the dimension of a Cartan torus coincides with the
rank of the Lie algebra, the Cartan torus is called {\it complete} for the group. 
In general a complete Cartan torus requires a special 
(orthogonal) basis. There are situations where there does exist a Cartan
subalgebra, but no complete Cartan torus.

\subsection{A Complete Cartan Torus for $\SU(n)$}

The Lie algebra $\log\SU(n)\cong\R^{n^2-1}$, $n\ge2$, 
in the defining complex $n$-dimensional representation has a basis 
consisting
of traceless and hermitian  generalized Pauli matrices
\begin{eq}{l}
\{\si(n)^a\}_{a=1}^{n^2-1},~\tr\si(n)^a=0,~~\si(n)^a=(\si(n)^a)^\star\cr
\end{eq}constructed inductively
from the proper Pauli matrices $\si(2)=\si$. The start
for $n\ge3$ is the embedded Lie subalgebra of $\SU(n-1)$ with
\begin{eq}{l}
\si(n+1)^a=
{\scriptsize\left( \begin{array}{c|c}
\si(n)^a&0\cr\hline0&0\cr\end{array}\right)},~~
a=1,\dots,n^2-1
\end{eq}The new off-diagonal matrices
for $a=n^2,\dots,(n+1)^2-2$ come in $(n-1)$ pairs 
with unit column vectors $e$ and their transposed
$e^T$ as illustrated in the first
step for the eight Gell-Mann matrices $\si(3)=\la$
\begin{eq}{rlrl}
\si(n+1)^a&={\scriptsize\left( \begin{array}{c|c}
\bl 0_n&e\cr\hline e^T&0\cr\end{array}\right)},&
\si(n+1)^{a+1}&={\scriptsize\left( \begin{array}{c|c}
\bl 0_n& -ie\cr\hline ie^T&0\cr\end{array}\right)}
\cr
\si(3)^4&={\scriptsize\left( \begin{array}{cc|c}
0&0&1\cr 0&0&0\cr\hline 1&0&0\cr\end{array}\right)},&
\si(3)^5&={\scriptsize\left( \begin{array}{cc|c}
0&0&-i\cr 0&0&0\cr\hline i&0&0\cr\end{array}\right)}\cr
\si(3)^6&={\scriptsize\left( \begin{array}{cc|c}
0&0&0\cr 0&0&1\cr\hline 0&1&0\cr\end{array}\right)},&
\si(3)^7&={\scriptsize\left( \begin{array}{cc|c}
0&0&0\cr 0&0&-i\cr\hline 0&i&0\cr\end{array}\right)}
\end{eq}The new diagonal matrix is defined by
\begin{eq}{l}
\si(n)^{n^2-1}={1\over\sqrt{{n\choose2}}}
{\scriptsize\left(\begin{array}{c|c}
\bl 1_{n-1}&0\cr\hline0&-(n-1)\cr\end{array}\right)}\cr
\end{eq}Therewith the  normalization  
is as for  the  proper Pauli matrices 
\begin{eq}{l}
\tr\si(n)^a\si(n)^b=2\de^{ab}
\end{eq}A Cartan subalgebra  $\log \U(1)^{n-1}$ is spanned by the 
diagonal matrices 
\begin{eq}{l}
\hbox{Cartan subalgebra basis: }
\{i\si(n)^{m^2-1}\mid m=2,3,\dots,n\}\cr
\end{eq}whose exponent gives a complete Cartan torus
of dimension $n-1$ (rank of $\log\SU(n)$).

The characteristic diagonal element with a nontrivial  
determinant generates the centrum of $\SU(n)$
and is renormalized to display  integer $\U(1)$-winding numbers  
in the diagonal
\begin{eq}{l}
\bl w_n=\sqrt{{n\choose2}}\si(n)^{n^2-1}
={\scriptsize\left(\begin{array}{c|c}
\bl 1_{n-1}&0\cr\hline0&-(n-1)\cr\end{array}\right)},~~\det\bl w_n=-(n-1)\cr
\U(1)_{n^2-1}=\{e^{i\al\bl w_n}\mid\al\in[0,2\pi[\}
\cr

e^{{2\pi i\over n}\bl w_n}=e^{{2\pi i\over n}}\bl1_n 
\in\U(\bl1_n)\cap\U(1)_{n^2-1}=\U(\bl1_n)\cap\SU(n)\cong\I(n)\cr
\end{eq}Here  $\U(\bl1_n)=\U(1)\bl1_n$
denotes the scalar phase group.

\subsection{A Complete Cartan Torus for Hyperisospin}

Hypercharge and isospin symmetry with 
central correlation, called hyperisospin
\begin{eq}{l}
{\U(1)\x\SU(2)\over\I(2)}\cong\U(2)\end{eq}has a Cartan subalgebra in the defining complex $2$-dimensional
representation
\begin{eq}{l}
\{i\al_0\bl1_2+i\al_3\si^3\mid\al_{0,3}\in[0,2\pi[\}\cong\R^2
\end{eq}Its exponent has as factors the scalar hypercharge and the 
3rd isospin component phase
group which, however, are no direct factors for a torus
\begin{eq}{l}
e^{i\al_0\bl1_2+i\al_3\si^3}\in\U(\bl1_2)\o\U(1)_3\cr
\end{eq}The parametrization has the following ambiguity
inherited from the common centrum
$\U(\bl1_2)\cap\SU(2)\cong\I(2)$
\begin{eq}{l}
(\al_0,\al_3)=(\pi,0)\cong(0,\pi),~~e^{i\pi\bl1_2}=e^{i\pi\si^3}=-\bl1_2
\in\I(2)\cong \U(\bl1_2)\cap\SU(2)
\end{eq}A Cartan torus of $\U(2)$ arises 
with a projector basis containing two orthogonal elements
\begin{eq}{l}
e^{i\al_+{\bl1_2+\si^3\over2}}
e^{i\al_-{\bl1_2-\si^3\over2}}
\in\U(1)_+\x\U(1)_-,~~\al_\pm=\al_0\pm\al_3\cr
\cl P_\pm(2)={\bl1_2\pm\si^3\over2},~~\cl P_+(2)\cl P_-(2)=0
\end{eq}

For the general case
\begin{eq}{l}
{\U(1)\x\SU(n)\over\I(n)}\cong\U(n),~~\U(\bl1_n)\cap\SU(n)\cong\I(n)
\end{eq}the exponent of a Cartan subalgebra 
in the defining complex $n$-dimensional
representation
\begin{eq}{l}
\{i\al_0\bl1_n+i{\SUM_{m=2}^n}\al_{m^2-1}\si(n)
^{m^2-1}\mid\al_{0,m}\in[0,2\pi[\}\cong\R^n
\end{eq}gives an Abelian group where the
scalar phase  factor is correlated with the centrum generating factor 
\begin{eq}{rll}
&\U(\bl1_n)\o \U(1)_{n^2-1}\x\U(1)^{n-2},&
e^{i\al\bl w_n}\in\U(1)_{n^2-1}\cr
\hbox{e.g. for }\U(3):&
\U(\bl1_3)\o \U(1)_8\x\U(1)_3,&e^{i\al\bl w_3}\in\U(1)_8,~~\bl w_3=\sqrt3\la^8
\end{eq}A Cartan torus comes with the appropriate
projectors $\cl P_\pm(n)$ and parameters $\al_\pm$
\begin{eq}{l}
\U(1)_+\x \U(1)_-\x\U(1)^{n-2},~~
e^{i\al_+\cl P_+(n)}
e^{i\al_-\cl P_+(n)}
\in\U(1)_+\x\U(1)_-\cr
\hbox{with }\left\{\begin{array}{ll}
\cl P_+(n)={(n-1)\bl 1_n+\bl w_n\over n},&\al_+=\al_0+
{\al_{n^2-1}\over\sqrt{{n\choose2}}}\cr
\cl P_-(n)={\bl 1_n-\bl w_n\over n},&\al_-=\al_0
-(n-1){\al_{n^2-1}\over\sqrt{{n\choose2}}}\cr
\end{array}\right.\cr
\cl P_+(n)\cl P_-(n)=0,~~
(\bl w_n)^2=(n-1)\bl1_n-(n-2)\bl w_n\cr
\end{eq}

For the groups $\U(n)$ with rank $n$ Lie algebras there exist
complete  Cartan tori.

\subsection{A Complete Cartan Torus for the Hydrogen Atom}

For the nonrelativistic hydrogen bound states 
an exponentiated Cartan subalgebra of $\log[\SU(2)\x\SU(2)]$
with basis $\{i\rvec\si\ox\bl1_2,\bl 1_2\ox i\rvec\tau\}$
in the defining quartet representation
\begin{eq}{l}
\hbox{Cartan algebra }\{i\al_3\si^3\ox \bl1_2
+\bl 1_2 \ox i\be_3 \tau^3\}\cong\R^2\cr
e^{i\al_3\si^3}\ox e^{i\be_3\tau^3}
={\scriptsize\pmatrix{
e^{i(\al_3+\be_3)}&0&0&0\cr 
0&e^{i(\al_3-\be_3)}&0&0\cr 
0&0&e^{-i(\al_3-\be_3)}&0\cr 
0&0&0&e^{-i(\al_3+\be_3)}\cr }}\in\U(1)_3\o\U(1)_3\cr
\hbox{parameters: }\{\al_3+\be_3,~\al_3-\be_3\}
\end{eq}leads to a complete Cartan torus via a basis of orthogonal generators 
$\cl L_\pm$ for coordinates $\ga_\pm$
\begin{eq}{l}
e^{i\al_3\si^3}\ox e^{i\be_3\tau^3}
=e^{i\ga_+\cl L^3_+}e^{i\ga_-\cl L^3_-}\in\U(1)_+\x \U(1)_-\cr
\cl L^3_\pm={\si^3\ox\bl1_2\pm\bl1_2\ox\tau^3\over2},~~
\cl L^3_+\cl L^3_-=0,~~\ga_\pm=\al_3\pm\be_3
\end{eq}$\cl L^3_+=\cl L^3$ is the 3rd component of the angular 
momenta $\log\SO(3)$, $\cl L^3_-\sim\cl F^3$ is proportional
to the 3rd component of the perihel vector.

In the general case two special groups, centrally correlatable
for dimensions with a common nontrivial factor
\begin{eq}{l}
{\SU(n)\x\SU(m)\over \I(k)},~~\I(k)\subnoteq\I(n)\cap\I(m),~~n,m,k\ge2 
\end{eq}the exponent of a Cartan Lie subalgebra is centrally correlated
by the $\U(1)$'s generated by 
$\bl w_n$ and $\bl w_m$
\begin{eq}{l}
\U(1)_{n^2-1}\o\U(1)_{m^2-1}\x\U(1)^{n+m-4}\cr
\hbox{$\U(1)_{n^2-1}\o\U(1)_{m^2-1}$-Lie algebra: }
\{i\al\bl w_n\ox \bl1_{m}
+\bl 1_n \ox i\be \bl w_{m}\}\cong\R^2\cr
e^{i\al\bl w_n}\ox e^{i\be\bl w_{m}}\in
\U(1)_{n^2-1}\o\U(1)_{m^2-1}\cr
\end{eq}In general, there arise
four parameters
\begin{eq}{l}
e^{i\al\bl w_n}\ox e^{i\be\bl w_{m}}
\cong{\scriptsize\pmatrix{
e^{i[\al+\be]}&0&0&0\cr 
0&e^{i[\al-(m-1)\be]}&0&0\cr 
0&0&e^{-i[(n-1)\al-\be]}&0\cr 
0&0&0&e^{-i[(n-1)\al+(m-1)\be]}\cr }}\cr
\hbox{parameters: }\{\al+\be,~\al-(m-1)\be,~(n-1)\al-\be,~
(n-1)\al-(m-1)\be\}
\end{eq}which only for the hydrogen symmetry with
$n=m=2$ allows an
orthogonal Cartan subalgebra basis leading to a complete Cartan torus.

\subsection{No Complete Cartan Torus for Hypercharge-Isospin-Color}

The internal interaction symmetry
$\U(2\x3)={\U(1)\x\SU(2)\x\SU(3)\over\I(2)\x\I(3)}$ 
 has  as defining complex 6-dimensional representation
for its Lie algebra with  rank 4 
\begin{eq}{l} 
\log[\U(1)\x\SU(2)\x\SU(3)]
=\{i\al_0\bl1_2\ox\bl1_3
+i\rvec \al\rvec\tau\ox\bl1_3
+\bl1_2\ox i\rvec\be\rvec\la\}\cong\R^{12}\cr
\hbox{Cartan subalgebra: }
\{i\al_0\bl1_2\ox\bl1_3
+i \al_3\tau^3\ox\bl1_3
+\bl1_2\ox i(\be_3\la^3+\be_8\la^8)\}\cong\R^{4}\cr
\end{eq}using three Pauli matrices $\rvec\tau$ (isospin)
and eight Gell-Mann matrices $\rvec\la$ (color).

The exponentiated Lie algebra has three correlated factors
generated with $\bl w_2=\si^3$ and $\bl w_3=\sqrt3\la^8$ 
\begin{eq}{rl}
&\U(\bl1_6)\o\U(1)_3\o\U(1)_8\x\U(1)\cr
e^{i\al_0\bl1_2\ox\bl1_3
+i \al_3\tau^3\ox\bl1_3
+\bl1_2\ox i \be_8\la^8}\in& \U(\bl1_6)\o\U(1)_3\o\U(1)_8\cr
\end{eq}The relevant parameter combinations in the
arising 4 phases 
\begin{eq}{l}
\hbox{parameters: }\{(\al_0\pm\al_3)+{\be_8\over\sqrt3},~~
(\al_0\pm\al_3)-{2\be_8\over\sqrt3}\}
\end{eq}cannot be disentangled with an
orthogonal basis for a representation of the  direct product
$\U(1)\x\U(1)\x\U(1)$.

There exists a complete Cartan torus $\U(1)_+\x\U(1)_-$ 
for hyperisospin $\U(2)$,
parametrized with $\{\al_0\pm\al_3\}$, 
there exists a complete Cartan torus $\U(1)_+\x\U(1)_-\x\U(1)_3$ 
 for hypercolor $\U(3)$,
parametrized with
 $\{\al_0+{\be_8\over\sqrt3},\al_0-{2\be_8\over\sqrt3},\be_3\}$. 
 However,  there does not exist a complete  Cartan torus $\U(1)^4$ for
faithful representations of the   internal $\U(2\x3)$-interaction symmetry.

\section{Eigenvector Bases for Correlated Groups}

A semisimple Lie algebra, and also $\log\U(n)$, allows - for any 
finite dimensional representation vector space -
a basis of eigenvectors for a Cartan subalgebra.
A Lie algebra representation involving also nontrivial nilpotent
transformations has not to have an eigenvector basis\cite{S89,BOE}.

Eigenvectors of a Cartan subalgebra have not to remain
eigenvectors for the exponentiated Cartan algebra. 
However, eigenvectors of a 
direct product of Abelian groups - of a Cartan torus in the compact case -  are needed
in the definition of particles (eigenstates).

In the following, `eigenvectors of a Lie algebra' 
and `eigenvectors of a Lie group'
are the acronyms for
`eigenvectors of a Cartan subalgebra' and `eigenvectors of a maximal 
direct product Abelian subgroup', in the case of  a compact Lie group 
of a maximal Cartan torus.
With the choice of an eigenvector basis (or of a Cartan subalgebra or
of a Cartan torus) the original full symmetry
seems to be broken.
However, the full symmetry remains in  the set with all possible eigenbases.
E.g. for spin $\SU(2)$
with a complete Cartan torus $\U(1)_3$: 
The 3rd direction choice to measure  spin eigenvalues
can be replaced equivalently by any direction.

Since a  correlation of two Lie  groups $G_1\x G_2$
via  a discrete centrum $C$ does not change
the Lie algebra
\begin{eq}{l}
\log{G_1\x G_2\over C}=\log[G_1\x G_2]=\log G_1\pl\log G_2
\end{eq}there can  arise the case where there exists an eigenvector basis
for the Lie algebra representation space 
which is no eigenvector basis
for the correlated group. This is the case
for compact groups without a complete Cartan torus, especially
for the internal interaction symmetry group.

\subsection{An Eigenvector Basis for $\U(n)$}

If  
a represented compact group has a 
complete Cartan torus there exists an eigenvector
basis  of the representation vector space - exemplified for $\U(n)$
and obviously true also for $\SU(n)$.

The diagonals of the $\log\U(n)$-Cartan
subalgebra basis in the defining representation
\begin{eq}{l}
\{i\bl 1_n,i\si(n)^{m^2-1}\mid m=2,3,\dots,n\}\cr
\end{eq}taken as columns  in the following 
 $(n\x n)$-scheme (between $||\dots||$)
display - up to $i$ - the eigenvalues
as components of the weights (in the  lines) with the eigenvectors 
$e^1={\scriptsize\pmatrix{1\cr0\cr\vdots\cr0\cr}}
,\dots,e^n={\scriptsize\pmatrix{0\cr\vdots\cr0\cr1\cr}}$

\begin{eq}{l}
 \begin{array}{c||c|ccccc||}
e^1:&1&1&{1\over\sqrt3}&{1\over\sqrt6}&\dots&{1\over\sqrt{{n\choose2}}}\cr
e^2:&1&-1&{1\over\sqrt3}&{1\over\sqrt6}&\dots&{1\over\sqrt{{n\choose2}}}\cr
e^3:&1&0&-{2\over\sqrt3}&{1\over\sqrt6}&\dots&{1\over\sqrt{{n\choose2}}}\cr
e^4:&1&0&0&-{3\over\sqrt6}&\dots&{1\over\sqrt{{n\choose2}}}\cr
.&.&&\dots&&\dots&\dots\cr
.&.& &\dots&&\dots&\dots\cr
e^n:&1&0&0&0&\dots&-{n-1\over\sqrt{{n\choose2}}}\cr\end{array}\cr 
\end{eq}E.g., for $\log \U(1)$
in the 
left upper $(1\x1)$-matrix, for $\log \U(2)$ in
the left upper $(2\x2)$-matrix, etc.
A geometrical aside, not really surprising
with the permutation group as symmetry group for
the fundamental $\SU(n)$-weights: 
Erasing the 1st column with the $1$'s for $\log\U(1)$, the
remaining $n$ lines ($\SU(n)$-weights with $n-1$ components
between $|\dots||$)
give the corners of a regular fundamental simplex 
(distance, triangle, tetraeder. etc.) centered at the origin of $\R^{n-1}$.

The  Lie algebra for the correlated group $\U(\bl1_n)\o\U(1)_{n^2-1}$ 
 has the 2-component weights
\begin{eq}{ll}
\hbox{for basis }
\{i\bl1_n,i\bl w_n\}: & \begin{array}{c||c|c||}
e^1:&1&1\cr
e^2:&1&1\cr
.&.&.\cr
e^{n-1}:&1&1\cr
e^n:&1&-(n-1)\cr
\end{array}\cr\cr 
\hbox{for projector basis }\{i\cl P_\pm(n)\}:& 
\begin{array}{c||c|c||}
e^1:&1&0\cr
e^2:&1&0\cr
.&.&.\cr
e^{n-1}:&1&0\cr
e^n:&0&1\cr\end{array}\cr 
\end{eq} 

Obviously, the eigenvectors keep their property for the complete Cartan torus
$\U(1)_+\x\U(1)_-\x\U(1)^{n-2}$, e.g. 
an eigenvector basis  for hyperisospin $\U(2)$  is given by
$\{e^1,e^2\}$ with
\begin{eq}{ll}
e^{i\al_+{\bl1_2+\si^3\over2}}
e^{i\al_-{\bl1_2-\si^3\over2}}e^1&=
e^{i\al_+{\bl1_2+\si^3\over2}}
e^1\cr
e^{i\al_+{\bl1_2+\si^3\over2}}
e^{i\al_-{\bl1_2-\si^3\over2}}e^2&=
e^{i\al_-{\bl1_2-\si^3\over2}}e^2
\end{eq}

\subsection{An Eigenvector Basis  for the Hydrogen Atom}

In the defining quartet $[{1\over2};{1\over2}]$ 
representation of the $\SO(4)$-invariant
bound state dynamics of the hydrogen atom the eigenvectors
of the Lie algebra as basis of the representation space $\C^2\ox\C^2$
\begin{eq}{l}
e^1\ox e^1
={\scriptsize\pmatrix{1\cr0\cr}}\ox {\scriptsize\pmatrix{1\cr0\cr}},~
e^1\ox e^2={\scriptsize\pmatrix{1\cr0\cr}}\ox {\scriptsize\pmatrix{0\cr1\cr}},~
 e^2\ox e^1,~ e^2\ox e^2
\end{eq}have the  eigenvalues
\begin{eq}{l}
\begin{array}{|c||c|c||c|c|}\hline
&\si^3\ox\bl1_2&\bl1_2\ox\tau^3&\cl L_+^3&\cl L_-^3\cr\hline\hline
e^1\ox e^1&+1&+1&+1&0\cr\hline
e^1\ox e^2&+1&-1&0&1\cr\hline
e^2\ox e^1&-1&+1&0&-1\cr\hline
e^2\ox e^2&-1&-1&-1&0\cr\hline
\end{array}~~\hbox{ with }
\cl L_\pm^3={\si^3\ox\bl1_2\pm \bl1_2\ox\tau^3\over 2}\cr
\end{eq}They remain eigenvectors of the 
correlated group $\SO(4)$ where they have to be characterized
by the orthogonal
basis $\cl L^3_+\cl L^3_-=0$.
The 3rd angular momentum $\cl L^3_+=\cl L^3$
component generates an axial rotation group 
$e^{i\ga_+\cl L^3_+}\in\SO(2)\subnoteq
\SO(3)$. The quartet comes as $\SO(3)$-triplet $(e^1\ox e^1,
{e^1\ox e^2+e^2\ox e^1\over\sqrt2},e^2\ox e^2)$
with $L=1$ and singlet ${e^1\ox e^2- e^2\ox e^1\over\sqrt 2}$ with $L=0$.
The 2nd basis element $\cl L^3_-={\cl F^3\over\sqrt{-2H}}$, 
generates also an Abelian subgroup $\SO(2)$, which, however,
is not a subgroup of another $\SO(3)$. Its The eigenvalue  is related to the 
number of radial knots $N$ in
the Schr\"odinger wave functions  $2J+1=L+1+N$ (not directly $N=L^3_-$!).

\subsection{No Eigenvector Basis for Hypercharge-Isospin-Color}

The defining representation of the Lie algebra $\log[\U(1)\x\SU(2)\x\SU(3)]$ 
on a
complex 6-dimensional space $\C^2\ox\C^3$ has
an eigenvector basis 
 with the eigenvalues read from the diagonal Pauli matrices
\begin{eq}{l}
\begin{array}{|c||c|c|c|c|}\hline
&Y\bl1_2\ox\bl1_3&\tau^3\ox\bl1_3&\bl1_2\ox\sqrt3\la^8
&\bl1_2\ox\la^3\cr\hline\hline
e^1\ox e^1&Y&+1&+1&+1\cr\hline
e^1\ox e^2&Y&+1&+1&-1\cr\hline
e^1\ox e^3&Y&+1&-2&0\cr\hline
e^2\ox e^1&Y&-1&+1&+1\cr\hline
e^2\ox e^2&Y&-1&+1&-1\cr\hline
e^2\ox e^3&Y&-1&-2&0\cr\hline
\end{array}
\end{eq}The normalization $Y\in\R$ will be discussed below.

Without a complete Cartan torus 
there do not exist eigenvector bases for the correlated group
$\U(2\x 3)$ in faithful  representations.

The subset of those $\U(2\x3)$-re\-pre\-sen\-ta\-tions
which are trivial either for  color or for isospin , 
i.e. the representations of hyperisospin $\U(2)$ or 
for hypercolor $\U(3)$, allow eigenvector bases
for $\U(2)$ and $\U(3)$ resp. They are obtained
from the corresponding fundamental representations
given by the antisymmetric cube or the antisymmetric square
of the defining $\U(2\x3)$-re\-pre\-sen\-ta\-tion 
which triples and doubles the hypercharge normalization.
Those product representations have 
the eigenvector bases
\begin{eq}{ll}
\begin{array}{l}
{\AND^3}u\in \U(2)\cr
\hbox{on }\C^2\cr
\hbox{with }3Y=1\end{array}&
\begin{array}{|c||c|c||c|c|}\hline
&3Y\bl1_2&\bl w_2&\cl P(2)_+&\cl P(2)_-\cr\hline\hline
e^1&3Y&+1&1&0\cr\hline
e^2&3Y&-1&0&1\cr\hline
\end{array}~~
\begin{array}{rl}
\bl w_2&=\tau^3\cr
\cl P(2)_\pm&={\bl1_2\pm\bl w_2\over2}\end{array}\cr\cr 
\begin{array}{l}
{\AND^2}u\in \U(3)\cr
\hbox{on }\C^3\cr 
\hbox{with }2Y=1\end{array}&
\begin{array}{|c||c|c|c||c|c|}\hline
&2Y\bl1_3&\bl w_3&\la^3
&\cl P(3)_+&\cl P(3)_-\cr\hline\hline
e^1&2Y&+1&+1&1&0\cr\hline
e^2&2Y&+1&-1&1&0\cr\hline
e^3&2Y&-2&0&0&1\cr\hline
\end{array}~~\begin{array}{rl}
\bl w_3&=\sqrt3 \la^8\cr
\cl P(3)_+&={2\bl1_3+\bl w_3\over3}\cr
\cl P(3)_-&={\bl1_3-\bl w_3\over3}\cr\end{array}
\end{eq}To obtain the projectors, the normalization $Y$ has to
fullfill  $3|Y|=1$ for $\U(2)$ and $2|Y|=1$ for $\U(3)$.

It is impossible to give an eigenvector
basis for the  internal group $\U(2\x3)$
in faithful representations, e.g. for the left-handed isodoublet color
triplet quark representation
$[{1\over6}||1;1,0]$. It is possible to give eigenvector bases for the 
reduced  internal groups $\U(2)$ or $\U(3)$,
 e.g. for the representations with the
 left-handed  isodoublet color singlet lepton 
or the right-handed  isosinglet color triplet  quarks resp.  
A quark confinement can be interpreted as
the decision with respect to a particle classification
for the complete Cartan torus 
$\U(1)_+\x\U(1)_-\subnoteq \U(2)$ for hyperisospin
and against the complete Cartan torus 
$\U(1)_+\x\U(1)_-\x\U(1)_3\subnoteq \U(3)$ for hypercolor.
  
With the reduction from $\U(2\x3)$ to hyperisospin $\U(2)$
the projector basis ${\bl1_2\pm\tau^3\over 2}$
generates the electromagnetic subgroup $\U(1)$
as one factor in the Cartan torus, let's take $\U(1)_+$.
With the choice of a projector basis  to
characterize eigenstates, 
no reduction from the interaction hyperisospin symmetry $\U(2)$
to the particle electromagnetic symmetry $\U(1)_+$ is enforced.

\section{The External-Internal Symmetry Correlation}

Also the external Lorentz group and the internal
hyperisospin-color group for the interaction symmetry transformations
are centrally correlated.

\subsection{Correlations by Defining Representations}

Correlations are implementable by specific representations, especially 
by defining representations.

A rank $r$ semisimple Lie algebra, e.g. 
$\log\SU(n)$ with $r=n-1$, has $r$-fundamental
representations, e.g. quark and antiquark representations $[1,0]$ and $[0,1]$
for $\log\SU(3)$, which are a basis - with respect
to totally symmetric tensor products - for all representations, e.g.
$[2C_1,2C_2]\sub {\OD^{2C_1}}[1,0]\ox{\OD^{2C_2}}[0,1]$.
A layer deeper are the  defining representations which are a subset of
the fundamental representations and allow, using also 
totally antisymmetric products, to construct all fundamental
representations, e.g. antitriplet from triplets $[0,1]\cong[1,0]\and[1,0]$.
If such a defining representation comes with a central correlation of the
represented groups, all its products will inherit this correlation.

The complex defining representation of $\SU(n)$ on $\C^{n}$ 
comes with a representation of 
the scalar  phase $\U(\bl1_n)$
\begin{eq}{rl}
\U(n)&\ni e^{i\al_0Y\bl1_n+i\rvec\al\rvec\si(n)}=
[Y||\underbrace{1,0,\dots,0}_{n-1~{\rm places}}]\cr
\hbox{e.g. }\U(2)&\ni e^{i\al_0Y\bl1_2+i\rvec\al\rvec\si}=
[Y||1]\cr
\U(3)&\ni e^{i\al_0Y\bl1_2+i\rvec\al\rvec\la}=
[Y||1,0]\cr
\end{eq}The correlation from the $\U(n)\cong {\U(1)\x\SU(n)\over\I(n)}$
representation
is inherited by all products, e.g. for the antisymmetric ones
with $n$-ality $k{\rm mod} n$
\begin{eq}{ll}
{\AND^k}[Y||1,0,\dots,0]
&=[kY||\underbrace{0,\dots,0,1,0,\dots,0}_{k{\rm -th ~place}}],~k=1,\dots,n-1\cr
{\AND^{n}}[Y||1,0,\dots,0]
&=[nY||0,\dots,0,0],~~nY\in\Z\cr
\end{eq}The $\U(1)$-representation for power $n$
has to come with an integer winding number, minimal for  $|nY|=1$.
For the examples above one obtains with minimal hypercharge $Y$
\begin{eq}{rl}
\U(2):&\left\{\begin{array}{rl}
[Y||1]\and [Y||1]&=[2Y||0]\in\U(1),~~|Y|={1\over 2}\cr
{\OD^{2T}}[Y||1]&=[2TY|2T|,~2T=0,1,\dots \cr\end{array}\right.\cr
\cr
\U(3):&\left\{\begin{array}{rl}
[Y||1,0]\and [Y||1,0]&=[2Y||0,1]\cr
{\AND^3}[Y||1,0]&=[3Y||0,0]\in\U(1),~~|Y|={1\over 3} \cr\end{array}\right.\cr
\end{eq}

In this way, if all interaction parametrizing fields of the standard 
model arise as representation products of one defining complex 6-dimensional 
representation on $\C^2\ox\C^3$, 
they display the central $\I(6)$-correlation
as given in ${\U(1)\x\SU(2)\x\SU(3)\over\I(6)}$,
e.g. for the fermion fields 
\begin{eq}{l}
u=[{1\over 6}||1;1,0]\then\left\{\begin{array}{llccl}
\hbox{quark isodoublet}&\bl q&\hbox{with}& u&=[{1\over 6}||1;1,0]\cr
\hbox{down antiquark isosinglet}&\bl d^\star&\hbox{with}& {\AND^2}u&=[{1\over 3}||0;0,1]\cr
\hbox{antilepton isodoublet}&\bl l^\star&\hbox{with}& {\AND^3}u&=[{1\over 2}||1;0,0]\cr
\hbox{up quark  isosinglet}&\bl u&\hbox{with}& {\AND^4}u&=[{2\over 3}||0;1,0]\cr
\hbox{lepton  isosinglet}&\bl e^\star&\hbox{with}& {\AND^6}u&=[1|0;0,0]\cr
\end{array}\right.
\end{eq}

Similarily, the defining quartet (2s and 2p states) representation 
$[{1\over2};{1\over2}]$ on $\C^2\ox\C^2$
for the bound states of the hydrogen atom
gives rise to all bound state representations
$[J;J]$ arising as direct summands in one of the totally symmetric products 
${\OD^N}[{1\over2};{1\over2}]$ acting on $\C^{{3+N\choose3}}$
\begin{eq}{l}
{\OD^N}[{1\over2};{1\over2}]=\left\{\begin{array}{lll}
{\PL_{J=0,1,\dots,{ N\over2}}}[J;J],
&N\hbox{ even}&\then 2J+1\hbox{ odd}\cr
{\PL_{J={1\over2},{3\over2},\dots, {N\over2}}}[J;J],
&N\hbox{ odd}&\then 2J+1\hbox{ even}\cr\end{array}\right.
\end{eq}All these 
representations inherit the $\I(2)$-correlation in ${\SU(2)\x\SU(2)\over\I(2)}$, 
nontrivial for even multiplicities $(2J+1)^2=4,16,\dots$.
 
\subsection{The Spin-Isospin-Correlation}

If the hadrons arise from  quark field products
they inherit the $\I(2)$-correlation from Lorentz $\SL(\C^2)$ and
isospin $\SU(2)$ in the fundamental representation on $\C^2\ox\C^2$, as seen
in the left handed Weyl doublet isodoublet 
color triplet quark representation ond $\C^2\ox\C^2\ox\C^3$,
faithful for the centrally correlated group
\begin{eq}{l}
{\SL(\C^2)\x\U(1)\x\SU(2)\x\SU(3)\over \I(2)\x\I(2)\x\I(3)}
\end{eq}and arising also
in the left handed  Weyl doublet isodoublet color singlet lepton representation
on $\C^2\ox\C^2$, faithful for 
\begin{eq}{l}
{\SL(\C^2)\x\U(1)\x\SU(2)\over \I(2)\x\I(2)}
\end{eq}One factor
 $\I(2)$ correlates spin in $\SL(\C^2)$ with isospin $\SU(2)$,
the other factor $\I(2)$ isospin $\SU(2)$ with hypercharge $\U(1)$.

A 3-dimensional Cartan subalgebra for a maximal compact 7-dimensional 
Lie subalgebra
for spin, hypercharge and isospin
\begin{eq}{l}
\R^7\cong \log[\SU(2)\x\U(1)\x\SU(2)]
\subnoteq \log[\SL(\C^2)\x\U(1)\x\SU(2)]\cong\R^{10}
\end{eq}in a
fundamental complex $4$-dimensional representation is given by
\begin{eq}{l}
\{i\om_3\si^3\ox\bl1_2+i\al_0\bl1_2\ox\bl1_2+\bl1_2\ox i\al_3\tau^3\}\cong\R^3
\end{eq}The exponent involves 4 parameters
\begin{eq}{l}
\U(1)_{\si^3}\o\U(\bl1_2)\o\U(1)_{\tau^3}=
\U(1)_{\si^3}\o[\U(1)_+\x\U(1)_-]\cr
\hbox{parameters: }
\{\pm\om_3+\al_0\pm\al_3\}=\{\pm\om_3+\al_+,\pm\om_3+\al_-\}
\end{eq}which prevent a complete $3$-dimensional Cartan torus for 
${\SU(2)\x\U(1)\x\SU(2)\over \I(2)\x\I(2)}$.
There exist complete $2$-dimensional  Cartan tori $\U(1)\x\U(1)$
for the centrally correlated two factor subgroups
\begin{eq}{l}
{\SU(2)\x\U(1)\over \I(2)}\cong\U(2),~~
{\SU(2)\x\SU(2)\over \I(2)}\cong\SO(4)
\end{eq}

Therefore, one has to decide 
with respect to eigenvector bases once more for a subgroup with a 2-dimensional
Cartan torus - the choice 
in the observed particles is $\U(2)$ with the scalar phase factor
the electromagnetic $\U(1)_+\subnoteq\U(2)$ from hyperisospin 
\begin{eq}{l}
\U(2)\cong {\SU(2)\x\U(1)_+\over \I(2)}\supnoteq
\U(1)_{\si^3}\o\U(1)_+,~~\hbox{ parameters: }
\{\pm\om_3+\al_+\}\cr
\hbox{Cartan torus: }
e^{i(\om_3+\al_+){\bl 1_2+\si^3\over2}\ox{\bl1_2+\tau^3\over2}}
e^{i(-\om_3+\al_+){\bl 1_2-\si^3\over2}\ox{\bl1_2+\tau^3\over2}}
\in\U(1)_{++}\x\U(1)_{-+}
\end{eq}The other hyperisospin circle  $\U(1)_-$
does not arise with eigenvectors. 
In the standard model
the  $\U(1)_-$ symmetry is
spontaneously broken via a degenerated 
ground state, implemented by the Higgs field $\bl\Phi$ in a
defining $\U(2)$-re\-pre\-sen\-ta\-tion
\begin{eq}{l}
\angle{\bl\Phi\ox \bl\Phi^\star}=
{\scriptsize\pmatrix{0\cr M\cr}}\ox (0,M)= 
{\scriptsize\pmatrix{0&0\cr 0&M^2\cr}}={\bl1_2-\tau^3\over 2} M^2
\end{eq}The group $\U(2)$
induces nontrivial isospin multiplicities
in the representation space (particles as translation eigenvectors) 
in contrast to the confined color.

\section{Summary}

The construction of eigenstates for the large homogeneous interaction
symmetry group
can be done in three steps ($\downarrow$), the first two ones
characterized by the choice of a maximal, but not complete Cartan torus 

{\scriptsize
\begin{eq}{c}

\begin{array}{|c|c|c|c|}\hline
&\hbox{\bf group}&\hbox{\bf defining field}&\hbox{\bf representation}\cr\hline\hline
\begin{array}{l}
\hbox{interaction}\cr
\hbox{operations}\end{array}
&{\SL(\C^2)\x\U(1)\x\SU(2)\x\SU(3)\over\I(2)\x\I(6)}
&\begin{array}{c}
\psi_{\al,i}\cr
\al=1,2\cr
i=1,2,3\end{array}
&\hbox{with }[{1\over 6}||1;1,0]\cr\hline\hline
\begin{array}{c}
\hbox{confinement}\cr
\hbox{of color $\SU(3)$}\end{array}
&\downarrow
&&\cr\hline

&{\SL(\C^2)\x\U(1)\x\SU(2)\over\I(2)\x\I(2)}
&\begin{array}{c}
({\AND^3}\psi)_\al\cr
\al=1,2\cr
\end{array}
&\hbox{with }[{1\over 2}||1]\cr\hline

\begin{array}{c}
\hbox{reduction}\cr
\U(2)\to \U(1)_+\cr
\hbox{to charge}\cr
\end{array}
&\downarrow
&&\cr\hline

&{\SL(\C^2)\x\U(1)_+\over\I(2)}
&\begin{array}{rl}
\bl p_+=&\bl\Phi^\al({\AND^3}\psi)_\al\cr
\bl n_0=&\bl\Phi^\star_\be\ep^{\be\al}({\AND^3}\psi)_\al\cr
\end{array}
&\hbox{with }[1],~[0]\cr\hline

\begin{array}{c}
\hbox{rest or}\cr
\hbox{momentum}\cr
\hbox{system}\cr
\end{array}
&\downarrow
&&\cr\hline\hline
\hbox{particles}
&\begin{array}{rl}m^2>0:&
\left\{\begin{array}{l}
{\SU(2)\x\U(1)_+\over\I(2)}\cr
\cong\U(2)\cr\end{array}\right.\cr
\cr
m=0:&\left\{\begin{array}{l}
{\U(1)_3\x\U(1)_+\over\I(2)}\cr
\cong\U(1)\x\U(1)\cr\end{array}\right.
\end{array}
& & \cr\hline

\end{array}\cr\cr
\hbox{\bf Particles as Eigenvectors for the Interaction Group}

\end{eq}
}

In the 3rd and 4th column only the internal representation properties
are given.

\end{document}